\documentclass[12pt]{article}

\usepackage[dvips]{graphicx}
\usepackage[english]{babel}

\usepackage{psfrag}
\usepackage{amsmath}
\usepackage{amsthm}
\usepackage{amsfonts}
\usepackage{amssymb}
\usepackage{slashed}

\catcode`\@=11

\newcommand{\tr}{\mathop{\rm tr}\nolimits}

\newcommand{\II}{\hbox{{1}\kern-.25em\hbox{l}}}

\newcommand{\lrD}{{D^{\hspace{-0.8em}
      \raisebox{0.8ex}{$\scriptstyle\leftrightarrow$}}}{}}
\newcommand{\lD}{{D^{\hspace{-0.8em}
      \raisebox{0.8ex}{$\scriptstyle\leftarrow$}}}{}}
\newcommand{\rD}{{D^{\hspace{-0.8em}
      \raisebox{0.8ex}{$\scriptstyle\rightarrow$}}}{}}

\newcommand{\lrn}{{\nabla^{\hspace{-0.8em}
      \raisebox{0.8ex}{$\scriptstyle\leftrightarrow$}}}{}}

\newcommand{\vev}[1]{\langle {#1}\rangle}

\begin{document}
\begin{titlepage}

\vskip3cm
\begin{center}

\textbf{\LARGE Twist two operators at finite volume.}

\vspace*{1.6cm}

{\large
A. Manashov$^{1,2}$
 and
A.~Sch\"afer$^1$
}

\vspace*{0.4cm}

{\sl
$^1$ Institut f\"ur Theoretische Physik, Universit\"at
                          Regensburg, D-93040 Regensburg, Germany \\
$^2$ Department of Theoretical Physics,  Sankt-Petersburg State University,
St.-Petersburg, Russia
}

\vspace*{0.8cm}

{\bf Abstract\\[10pt]}
\parbox[t]{\textwidth}{
{We calculate the volume corrections   to the two-pion matrix element of twist two operators.}}

\vspace{1cm}

\end{center}

\end{titlepage}

\noindent {\bf 1.}\hskip 5mm Generalized parton distributions (GPDs) provide a unified
parametrization for nonforward hadronic matrix elements \cite{r1,r2,r3,r4}.
Understanding GPDs in detail is therefore tantamount to understanding in large parts the
internal structure of hadrons. The comprehensive review of  experimental programs as well
as theoretical works can be found in Refs.~\cite{r5,r6,r7}.

As a rule GPDs enter into physical observables only within convolutions.
These are relatively simple at leading order in $\alpha_s$
 but become increasingly complex at higher orders, see
e.g.~\cite{NNLO}.
It makes
the extraction of GPDs from experiment  a highly non-trivial task.
There are much hopes that the  lattice QCD can be  helpful for solving
this problem.
There already exists a number of lattice
results for moments of GPDs, see \cite{Edwards:2006qx,Gockeler:2005vz,Brommel06}
and references therein.

The lattice evaluation  of the moments of GPDs  is based on  calculation of  the form
factors of local operators, very similar to the case of the
usual electromagnetic form factors \cite{FF1}. Unfortunately, at present the lattice calculations
with dynamical quarks can be done only for unphysically large quark masses and thus pion.
The mass of the pion affects strongly the spatial extent of the nucleon and hence its
form factors. Since the calculations are restricted to  finite size lattice
additional corresponding corrections have also to be taken into account.

Both problems~--~calculation of the mass and volume corrections~--~can be addressed
in the framework of  chiral perturbation theory (ChPT)~\cite{Weinberg,GL,GL1,GLfin}.
These corrections  were already calculated
 for  a number of physical observables (masses, decay
constants, etc.). In particular,
the mass corrections to GPDs (pion and nucleon) have been calculated recently in
Refs.~\cite{KP-I,DMS05,DMS06,DMS06a,ACK06}.
The  calculation of the volume corrections  in the framework of ChPT was initiated
by the paper of Gasser and Leutwyler~\cite{GLfin}. Recently, the problem became rather
actual due to the needs of   lattice QCD.
A review of  the present state of the art and references can be found in Ref.~\cite{BM06}
(see also Ref.~\cite{CDS06} where  
the pion GPDs were calculated in partially-quenched ChPT at finite volume).

On a lattice the overall spatial symmetry is reduced from the usual Lorentz group to the
hypercubic group. As a result the restrictions imposed by this symmetry on the form of
matrix elements of twist--two operators are less stringent than in the continuum limit.
The number of independent formfactors increases very fast with the spin of an operator,
and the reduced symmetry   also results in additional mixing among
operators~\cite{Get96}. All this makes the  extraction of the GPD moments from lattice
data a rather complicated problem. Therefore, the knowledge of volume corrections to the
matrix elements of twist-two operators would be very helpful for determining  GPDs
moments on the lattice. In the present Letter we report the calculation of the volume
corrections to  the lowest moments of vector and tensor GPDs for the pion.

\vskip 5mm

\noindent
{\bf 2.}\hskip 5mm
To set the notations  we recall  the main ingredients of  chiral perturbation
theory~\cite{Weinberg,GL}.
ChPT  is an effective theory describing pion interactions
 in the limit $q, m_\pi\ll \Lambda_\chi$, where $q$
is a generic momentum and $\Lambda_\chi$ is the scale of  chiral symmetry breaking. We
will use the standard ${\cal O}(q^n)$ power-counting of ChPT. The Goldstone boson fields
are collected in the matrix-valued field $U$. The leading-order pion Lagrangian reads
\cite{GL}
\begin{align}	
\label{LO}
{\mathcal L}_{\pi\pi}^{(2)}=\frac{F^2}{4}
\tr\left(D_\mu U D^\mu U^\dagger+\chi^\dagger U+\chi U^\dagger\right)\,,
\end{align}	
where  the covariant derivative $D_\mu$ and the
external field tensor $\chi$ are defined by
\begin{align}	
D_\mu U= \partial_\mu U-i{r}_\mu\,U+iU\,{l}_\mu \,, &&
\chi   = 2B_0\left({s}+i\,{p}\right) \,.
\end{align}	
Here ${s}$ and ${q}$ denote external scalar and pseudoscalar fields,
which count as quantities of order ${\cal O}(q^2)$.  Furthermore,
${r}_\mu$ and ${l}_\mu$ are external right- and left-handed vector
fields with intrinsic chiral power ${\cal O}(q)$.  The two
leading-order parameters appearing in (\ref{LO}) are the pion decay
constant $F$ (normalized to $F \approx 92$~MeV) and the two-flavor
chiral condensate $B_0=- \langle 0|\bar{q}q|0 \rangle /F^2$, both
evaluated in the chiral limit \cite{GL}.  Throughout this work we use
the non-linear representation for the pion fields $\pi^a$,
\begin{eqnarray}\label{U}
U=\exp\{i\tau^{a}\pi^{a}\varphi((\pi/F)^2)/F \} \,,
\end{eqnarray}
where $\tau^a$ are the Pauli matrices.  The choice of the function $\varphi$ is more or
less arbitrary. We find it convenient to fix it  by the condition $d\mu(U)=\prod_{a=1}^3d\pi^a$,
where $d\mu(U)$ is the invariant measure on the  group $SU(2)$. This requirement results
in the following equation
\begin{align}\label{U-f}	
\varphi(x^2)-\frac{1}{2x}\sin \left(2x\,\varphi(x^2)\right)=\frac23 x^2\,, &&
\varphi(x^2)=1+\frac1{15}x^2+\frac{2}{175} x^4+O(x^6)\,.
\end{align}	
The leading order pion Lagrangian in the parametrization~(\ref{U}),~(\ref{U-f}) reads
\begin{align}	
\mathcal{L}^{(2)}_{\pi\pi}=
\frac12\left\{(\partial\pi)^2-m^2 \pi^2\right\}+\frac{1}{10F^2}\left\{
3(\pi\partial\pi)^2-\pi^2(\partial\pi)^2\right\}-\frac{m^2}{40F^2} (\pi^2)^2+O(\pi^6)\,.
\end{align}
The specific feature of this parametrization is the absence of pion field renormalization at
one loop, $Z_\pi^{one-loop}=1$.
\vskip 2mm

The chiral perturbation theory for pions in a finite volume was formulated in Ref.~\cite{GLfin}.
It was argued that imposing  periodic boundary conditions on the fields,
$$
\pi^a(x_0,x_1+L,x_2,x_3)=\pi^a(x_0,x_1,x_2+L,x_3)=\pi^a(x_0,x_1,x_2,x_3+L)
=\pi^a(x_0,x_i),
$$
does not affect
the  Effective Lagrangian~\eqref{LO}~\footnote[2]{We will not require a periodicity in  time
direction.}.
The only effect is a modification of
the pion propagator according to
\begin{align}	\label{pr-fin}
G^{ab}(x,x')=\sum_{\vec{n}} G_\infty^{ab}(x_0-x'_0,\vec{x}-\vec{x}'+\vec{n}L)\,,
\end{align}	
where the sum is taken over integer $\vec{n}$ and $G_\infty$ is  the
pion propagator in infinite volume
\begin{align}\label{inf-v-p}
G^{ab}_\infty(x)=\delta^{ab}\frac1i\int \frac{d^4p}{(2\pi)^4}\frac{e^{ipx}}{p^2-m^2+i0}\,.
\end{align}
The imposed boundary conditions reduce the Lorentz symmetry to
$R_0\times H(3)$, where $R_0$ is the reflection $x_0\to -x_0$ and $H(3)$ is the
cubic group.

\vskip 5mm

\noindent
{\bf 3.}\hskip 5mm
The first moment of the isovector GPD for a pion, (see Ref.~\cite{DMS05} for a definition)
is related to the matrix element of
the isovector vector operator $Q_\mu^a=\bar q\tau^a\gamma_\mu q$
\begin{align}\label{iso-first}
\vev{\pi^{b}(p')|Q_\mu^c|\pi^{a}(p)}=i\epsilon^{abc} A_{1,0}^{I=1}(t) P_\mu\,,&&
A_{1,0}^{I=1}(t)=\int_{-1}^{1}dx H^{I=1}(x,\xi,t)\,.
\end{align}
Throughout this Letter we  use the standard notations for the kinematical variables
 \begin{align}	
P=\frac12(p+p')\,,&&\Delta=p'-p\,, && t=\Delta^2\,, && \xi=-\frac{\Delta_+}{2P_+}\,,
\end{align}	
where $\Delta_+(P_+)$ is the  light-cone projection of the vector $\Delta(P)$.

\begin{figure}[t]
\psfrag{a}[c][c][0.8]{$P-\frac{\Delta}{2}$}
\psfrag{c}[c][c][0.8]{$P+\frac{\Delta}{2}$}
\psfrag{b}[c][c][0.8]{$l+\frac{\Delta}{2}$}
\psfrag{d}[c][c][0.8]{$l-\frac{\Delta}{2}$}
\psfrag{f}[c][c][0.8]{$l$}
\psfrag{aa}[b][b][1.0]{a}
\psfrag{bb}[b][b][1.0]{b}
\centerline{\includegraphics[width=10cm]{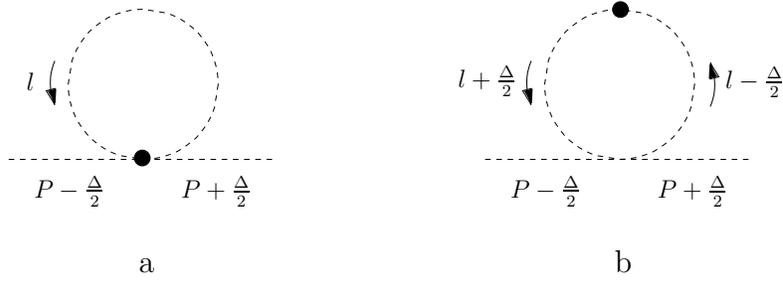}}
\caption{\label{fig1} One-loop graphs contributing to two-pion matrix
  elements of the pion operator.  The operator insertion is denoted
  by a black blob.}
\end{figure}

In ChPT the operator $Q_\mu^a$ takes the following form
(we will follow the notations of Ref.~\cite{DMS05})
\begin{align}	
O_\mu^a=-\frac{i}{2}F^2\left(L_\mu^a+R_\mu^a\right)+O(q^3)\,,
\end{align}	
where
$L_\mu=U^\dagger\partial_\mu U$,
$ R_\mu=U\partial_\mu U^\dagger$
and $L_\mu=L_\mu^a\tau^a$ and the same for $R_\mu$. Calculating the corrections to the matrix
element of this operator in ChPT one determines the formfactor $A_{1,0}^{I=1}(t)$.
The  infinite volume result for   $A_{1,0}^{I=1}(t)$  at order $O(q^2)$ reads~\cite{GL}
\begin{align}	\label{isom}
A_{1,0}(t)=2\,\Big(1+F_V(t)
-l^r_6(\mu)t/F^2\Big)\,,
\end{align}	
where $l^r_6(\mu)$ is a low energy constant (LEC),
\begin{align}	
F_V(t)=\frac{1}{6\Lambda_\chi^2}\left[(t-4m^2)J(t)-t\left(\frac13+\log
\frac{m^2}{\mu^2}\right)\right]\,,&&J(t)=2+\sigma\log\frac{\sigma-1}{\sigma+1}\,,
\end{align}	
$\Lambda_\chi=4\pi F$ and   $ \sigma=\sqrt{1-4m^2/t}$.  To obtain this result one has
to calculate the diagrams shown in Fig.~\ref{fig1}.
\vskip 5mm

To calculate the matrix element~(\ref{iso-first}) in a finite volume one has to evaluate
the same diagrams,   Fig.~\ref{fig1}, but with the propagator~(\ref{pr-fin}).
We
notice that the field renormalization is absent both in a infinite and finite volume.
Calculating the tadpole diagram in~ Fig.~\ref{fig1} (a) one notices that the term with
$\vec{n}=0$ in the sum in Eq.~(\ref{pr-fin}) gives rise to the infinite volume result,
while the remaining terms give corrections which vanishes at $L\to \infty$.
It is convenient to carry out a similar rearrangement for the diagram (b) in
Fig.~\ref{fig1} also. Schematically it amounts to the following transformations~\cite{Luscher85}
\begin{align}\label{luscher}	
\int_V d^d x& e^{ip x}G^2(x-y)=
\sum_{\vec{n}}\int_V d^d x e^{ip x}G_\infty(x-y+\vec{n}L)G(x-y)
=\int d^d x e^{ip x} G_\infty(x-y)G(x-y)\nonumber\\
&=\int d^d x e^{ip x} G_\infty(x-y)G_\infty(x-y)+{\sum_{\vec{n}}}'
\int d^d x e^{ip x} G_\infty(x-y)G_\infty(x-y+\vec{n}L)
\,.
\end{align}
Again the first term in the last line of Eq.~\eqref{luscher} represents the infinite
volume contribution
while the last one gives $1/L$ corrections. As usually the sign
${\sum_{\vec{n}}}'$ means that the term with
$\vec{n}=0$ in the sum  has to be omitted. The integration in last line of Eq.~(\ref{luscher})
goes over infinite space and involves infinite volume pion propagators~(\ref{inf-v-p}).
We notice also that each term in the sum~(\ref{luscher}) is UV finite.
One also has to take into account that  due to the imposed boundary condition all spatial momenta
are quantized, $\vec{q}=\left(2\pi/L\right) \vec{n}$, where $\vec{n}=(n_1,n_2,n_3)$ is a vector with
integer components. Then,
after lengthy but straightforward calculations
one obtains the following answer for  $1/L$ corrections to the matrix element~(\ref{iso-first})
\begin{align}\label{iso1L}	
\vev{\pi^b(p')|O^{c,\mu}|\pi^a(p)}_{1/L}=2i\epsilon^{abc}\Big(
P^\mu\, (A-B)-{\widetilde P}^\mu C+\Delta^\mu (\vec{P}\vec{S})\Big)\,.
\end{align}	
Here $\widetilde P=(0,\vec{P})$ and the formfactors $A,B,C,S^j$ are given by the following expressions
\begin{subequations}\label{ABC}
\begin{align}	
A=&\frac{2}{L\Lambda_\chi^2}{\sum_{n\in \mathbb{Z}^3}}'\int_{-1}^1d\eta\,
M_n(\Delta,\eta)\,\mathbb{K}_1(m(\eta),\vec{n})\,,
\\
B=&\frac{4}{L\Lambda_\chi^2}{\sum_{n\in \mathbb{Z}^3}}'\mathbb{K}_1(m,\vec{n})\,,
\\
C=&\frac{2}{3\Lambda_\chi^2}{\sum_{n\in \mathbb{Z}^3}}' \vec{n}^2
\int_{-1}^1d\eta\, M_n(\Delta,\eta)\,\mathbb{K}_2(m(\eta),\vec{n})\,,
\\
\label{Sj}
S^j=&\frac{1}{\Lambda_\chi^2} {\sum_{n\in \mathbb{Z}^3}}' {n_j}
\int_{-1}^1d\eta\,\eta\, M_{n}^j(\Delta,\eta)\,\mathbb{K}_1(m(\eta),\vec{n})\,,
\end{align}	
\end{subequations}
where
$
m^2(\eta)=m^2-\frac14(1-\eta^2)\Delta^2
$\,,
$\Delta^\mu=(\Delta_0,\vec{\Delta})$ and we put $\vec{\Delta}\equiv\dfrac{2\pi}{L} \vec{k}$.	
The function $M_n,M_n^j$ and $\mathbb{K}$ are defined as follows
\begin{align}	
M_n(\Delta,\eta)=&\prod_{m=1}^3 (-1)^{n_m k_m}\cos(\pi\eta\, n_mk_m)\,,&&
M_n^j(\Delta,\eta)=\frac{\sin(\pi\eta\, n_jk_j)}{\cos(\pi\eta\, n_jk_j)} M_n(\Delta,\eta)\\
\intertext{and}
\mathbb{K}_i(m,\vec{n})=&\left(\frac{m}{|\vec{n}|}\right)^i\, K_i(m|\vec{n}|L)\,,
\end{align}
where	$K_i(x)$ is the modified Bessel function of the second kind. The sums
in~(\ref{ABC}) converge rapidly since $K_i(x)\sim x^{-1/2}e^{-x}$ for large $x$.
Since the Lorentz symmetry does not hold for $1/L$ corrections the matrix
element~(\ref{iso1L}) is parameterized by three formfactors instead of one in an infinite
volume limit. Let us remark that the difference $A-B$ vanishes in the limit $\Delta\to0$
which ensures that the temporal component, $\mu=0$, of  matrix element~(\ref{iso-first})
does not receive
corrections in this limit.

\vskip 5mm

\noindent
{\bf 4.} \hskip 5mm
The first moment of the isoscalar pion  GPD, $H^{I=0}(x)$, is parameterized by two
formfactors,
$A_{2,0}^{I=0}$ and $A_{2,2}^{I=0}$,
\begin{align}\label{}
\int_{-1}^1 dx H^{I=0}(x,\xi,t)=A_{2,0}^{I=0}(t)+4\xi^2 A_{2,2}^{I=0}(t)\,.	
\end{align}
They enter the matrix elements of the isoscalar operator,
\begin{align}	\label{s2}
&Q_{\mu\nu}^s=
\underset{\mu\nu}{\mathrm{S}}
\left(\bar q\,\gamma_\mu
i\lrD_\nu \,q-\dfrac14 g_{\mu\nu} \bar q i\slashed{\lrD} q\right)\,,\\
\label{me2}
&\vev{\pi^b(p')|Q_2^s(z)|\pi^a(p)}=\delta^{ab} \Big(A_{20}^{I=0}(P\cdot z)^2+A_{22}^{I=0}
 (\Delta\cdot z)^2\Big)\,,
\end{align}	
where $z$ is auxiliary light-like vector, 
${\lrD}_\mu=\dfrac12\left(\rD_\mu-\lD_\mu\right) $,
$\underset{\mu\nu}{\mathrm{S}}\,Q_{\mu\nu}=
\frac12(Q_{\mu\nu}+Q_{\nu\mu})$ and we set $Q(z)=z_\mu z_\nu Q^{\mu\nu}$.
We also notice, that  throughout the   paper  the scalar product $(a\cdot b)$ implies  the Minkowsky
scalar product.

In ChPT the
operator~(\ref{s2}) reads
\begin{align}	
Q_{\mu\nu}^s=F^2\tilde a_{20}\left(L_{\mu}^a L_{\nu}^a+R_{\mu}^a
R_{\nu}^a-\frac{g_{\mu\nu}}4\left[L_\rho^a L^{\rho,a}+R_\rho^a R^{\rho,a}\right]\right)\,.
\end{align}	
In an infinite volume the formfactor $A_{20}$ does not receive nonanalytic corrections at
order $O(q^2)$, while the result for the formfactor $A_{22}$ reads~\cite{DL91}
\begin{align}	
A_{22}^{I=0}(t)=A_{22}^{(0)}\left(1+\frac{m^2-2t^2}{3\Lambda_\chi^2}\left[\log\frac{m^2}{\mu^2}+
\frac43-\frac{t+2m^2}{t}J(t)\right]\right)+A_{22}^{(1,m)}m^2+A_{22}^{(1,t)}t\,.
\end{align}	
We recall also that $A_{22}^{(0)}=-A_{20}^{(0)}/4=\tilde a_{20}$.

The volume  corrections  to the matrix element~(\ref{me2}) comes from diagram (b) in
Fig.~(\ref{fig1}) only,
the contribution due to the tadpole diagram,~Fig.~\ref{fig1}(a), vanishes.
The calculation gives the following result for the matrix element
\begin{align}	
\vev{\pi^b(p')|Q_2^s(z)|\pi^a(p)}_{1/L}=&\tilde a_{2,0}\left(t-\frac{m^2}{2}\right)\,
 T(z)\,,\\[2mm]
T(z)=&	
\left[(\Delta\cdot  z)^2 D - (\Delta\cdot  z) (\vec{z}\vec{S})+\frac12\vec{z}^2 C)\right]
\,,
\end{align}	
where $T(z)=z_\mu z_\nu T^{\mu\nu}$ and
\begin{align}	
D=&\frac1{4\Lambda_\chi^2}{\sum_{\vec{n}\in \mathbb{Z}^3}}'\int_{-1}^{1}d\eta\,
(1-\eta^2)\,
\,M_n(\Delta,\eta)\,\mathbb{K}_0(m(\eta),\vec{n})\,.
\end{align}	

It is interesting to notice  that both the infinite and finite volume corrections  vanish at $2t=m^2$.
To make the structure of corrections
 more transparent we write down  the tensor $T^{\mu\nu}$ in components
\begin{subequations}
\begin{align}	
T^{00}=&\left((\Delta^0)^2-\frac14 \Delta^2\right)D+\frac14(\vec{\Delta}\vec{S})+\frac34
C\,,\\
T^{0i}=&T^{i0}=\Delta^0\left(\Delta^i D-\frac12 S^i\right)\,,\\
T^{ik}=&\Delta^i\Delta^k D+\frac12\left(\Delta^i S^k+\Delta^k S^i\right)
+\frac{\delta_{ik}}{4}\left(\Delta^2-(\vec{\Delta}\vec{S})+\frac14 C\right)\,.
\end{align}	
\end{subequations}
Thus  parametrization of the volume corrections for an isoscalar operator at one-loop level
require only one
additional  function, $D$.

\vskip 5mm

\noindent
{\bf 5.}\hskip 5mm
The lowest moments of chiral odd GPDs, $E^{I}_{T\pi}$,
are related to the matrix elements of the operators
$Q_{\lambda\mu}^a=\bar q \sigma_{\lambda\mu} \tau^a q$ and
$Q^{(s)}_{\lambda\mu,\rho}=\underset{\lambda\mu}{\mathrm{A}}\,\underset{\mu\rho}{\mathrm{S}}\,
\bar q\sigma_{\lambda\mu} i\lrD_\rho q $~\cite{DMS06}. The symbols
$\underset{\mu\rho}{\mathrm{S}}$ and
$\underset{\lambda\mu}{\mathrm{A}} $
denote (anti)symmetrization in the corresponding indices. The pion matrix
elements of these operators are parameterized as follows
\begin{align}\label{mecho}
\vev{\pi^c(p')| Q_{\lambda\mu}^a |\pi^b(p)}=&i\epsilon^{abc} R_{\lambda\mu}
B^\pi_{T1,0}(t)\,,\\[2mm]
\vev{\pi^c(p')| Q_{\lambda\mu,\rho}^{(s)} |\pi^b(p)}=&\delta^{bc}
\underset{\lambda\mu}{\mathrm{A}}\,\underset{\mu\rho}{\mathrm{S}}
 R_{\lambda\mu} P_\rho
B^\pi_{T2,0}(t)\,,
\end{align}
where $R_{\lambda\mu}=(P_\lambda \Delta_\mu-\Delta_\lambda P_\mu)/m_\pi$. The formfactors
$B^\pi_{T,n0}$ are equal to the moments of the chirally odd GPD
$E^{I}_{T\pi}$, $\left(B^\pi_{T,n0}(t)=\int_{-1}^1 dx x^{n-1} E_{T,\pi}(x,\xi,t)\right)$.
In  ChPT these operators take the form (we follow the notations of Ref.~\cite{DMS06a})
\begin{align}	\label{odd}
Q^{a}_{\lambda\mu}=&\frac1{8}F^2 b_{T1,0} \tr \Big(u^\dagger \tau^a u^\dagger+
u\tau^a u\Big) [u^\lambda,u^\mu]\,, \\
\label{odd-s}
Q^{(s)}_{\lambda\mu,\rho}=&\frac1{8}F^2 b_{T2,0} \tr \Big(u^2+(u^\dagger)^2\Big)
[u^\lambda(2i\lrn^\rho)
u^\mu-(\lambda\leftrightarrow\mu)]\,,
\end{align}	
where $u^2=U$,
\begin{align*}	
u_\mu=i\left(u^\dagger \partial_\mu u -u\partial_\mu u^\dagger\right),&&
\nabla_\mu X=\partial_\mu X + [\Gamma_\mu, X]\,,&&
\Gamma_\mu=\frac12\left(u^\dagger \partial_\mu u +u\partial_\mu u^\dagger\right)\,.
\end{align*}	
In Eq.~(\ref{odd-s}) it is implied that one has to subtract all traces and
implement the indicated (anti)symmetrization
\begin{align}\label{symasym}	
T^{\lambda\mu,\rho}\to \frac14\left(
2T^{\lambda\mu,\rho}+T^{\lambda\rho,\mu}-T^{\mu\rho,\lambda}-g^{\mu\rho}{T^{\lambda\sigma}}_\sigma+
g^{\lambda\rho}{T^{\mu\sigma}}_\sigma
\right)\,.
\end{align}	
The formfactors $B^\pi_{T,n0}(t)$ at  order $O(q^2)$ (in an infinite volume) were
calculated in~\cite{DMS06a}
\begin{align}	
B^\pi_{T,10}(t)=&B^{(0)\pi}_{T,10}
\left(1+\frac{m^2}{2\Lambda_\chi^2}\,\log\frac{m^2}{\mu^2}+F_V(t,m^2)\right)
+\ldots\,,\\
B^\pi_{T,10}(t)=&B^{(0)\pi}_{T,20}
\left(1-\frac{3m^2}{2\Lambda_\chi^2}\,\log\frac{m^2}{\mu^2}\right)+\ldots\,,
\end{align}	
where the dots stand for analytic terms and $B^{(0)\pi}_{T,10}=m_\pi b_{T1,0}$,
$B^{(0)\pi}_{T,20}=-2m_\pi b_{T2,0}$.

Calculating the diagrams in Fig.~\ref{fig1} with the insertion of the operators
$Q^a_{\lambda\mu}$ and $Q^{(s)}_{\lambda\mu,\rho}$ one obtains the following result
for the volume corrections to the matrix elements
\begin{align}	\label{odd-iso}
\vev{\pi^c(p')|Q^{a}_{\lambda\mu}|\pi^b(p)}_{1/L}=&
i\epsilon^{abc}\,\frac{B^{(0)\pi}_{T,10}}{m_\pi}
\Biggl\{\left(P^\lambda \Delta^\mu-\Delta^\lambda P^\mu\right) \left(A-\frac{B}{2}\right)
-(\widetilde P^\lambda \Delta^\mu-\widetilde P^\mu \Delta^\lambda)\,C\Biggr\}\,.
\end{align}	
We recall that $\widetilde P^0=0$ and $\widetilde P^i=P^i$.
Let us notice that  $1/L$ correction to the matrix element of the chirally odd isovector
operator~(\ref{odd-iso}) involves the functions $A,B$ and $C$ which enter the volume correction
to the chirally even isovector operator~(\ref{iso1L}).

In its turn the volume correction to the chirally odd isoscalar operator
\begin{align}	\label{q2sc}
&\vev{\pi^b(p')|Q^{(s)}_{\lambda\mu,\rho}|\pi^a(p)}_{1/L}={\delta^{ab}} b_{20}\Big\{
-\frac{3}{2}P^\rho\,\left(P^\lambda \Delta^\mu-\Delta^\lambda P^\mu\right)\, B\nonumber\\
&\hspace{2cm}+\delta^{\rho0}\left(\delta^{\lambda0} \Delta^\mu-\delta^{\mu0} \Delta^\lambda\right)
\left[E-\left(t-\frac{m^2}{2}\right)\,C\right] +\left(t-\frac{m^2}{2}\right)\Delta^\rho\left(S^\lambda
\Delta^\mu-\Delta^\lambda S^\mu\right)\,,
\end{align}	
involves one new function $E$,
\begin{align}	
E=&\frac{4}{3L\Lambda_\chi^2}{\sum_{\vec{n}}}'\,\vec{n}^2\,\mathbb{K}_3(m,\vec{n})\,.
\end{align}
We also introduced the notations $S^\mu=(0,\vec{S})$, where $\vec{S}$ is defined by
Eq.~(\ref{Sj}).
Of course, the  matrix element~\eqref{q2sc} has to by symmetrized according to~\eqref{symasym}.

\vskip 5mm
\noindent
{\bf 6.} \hskip 5mm In this Letter we have presented the results of  the calculation of volume
corrections to the matrix elements of the pion operators of lowest spin.
We found that the
$1/L$ corrections to the four  operators we considered involve  (due to  Lorentz symmetry
breaking)  11 different kinematical structures with  coefficients parameterized by
six functions ($A,B,C,D,E,S^j$).  Our results should be helpful to improve the procedure of the
extracting of GPD moments from lattice data.

\vskip 5mm
\noindent
{\bf Acknowledgments}

\vskip 5mm

This work was supported
by the Helmholtz Association, contract number VH-NG-004.

\end{document}